\documentclass[pre,groupaddress,twocolumn,showpacs,amsmath,aps]{revtex4}
\usepackage{epsfig,graphicx,hyperref}
\usepackage{dcolumn}
\usepackage{bm}
\newcommand{\be}{\begin{equation}}
\newcommand{\ee}{\end{equation}}
\newcommand{\ba}{\begin{eqnarray}}
\newcommand{\ea}{\end{eqnarray}}
\renewcommand{\phi}{\varphi}

\begin{document}

\title{The role of attractive forces in viscous liquids}

\author{Ludovic Berthier} 
\affiliation{Laboratoire Charles Coulomb, UMR 5221, CNRS and Universit\'e
Montpellier 2, Montpellier, France}

\author{Gilles Tarjus} 
\affiliation{LPTMC, UMR 7600, CNRS and Universit\'e Pierre et Marie Curie,
4 Place Jussieu, 75252 Paris Cedex 05, France}

\date{\today}

\begin{abstract} 
We present evidence from computer simulation that the
slowdown of relaxation of a standard Lennard-Jones glass-forming liquid and
that of its reduction to a model with truncated pair potentials without
attractive tails is quantitatively and qualitatively different in the
viscous regime. The pair structure of the two models is however very
similar. This finding, which appears to contradict the common view that 
the physics of dense liquids is dominated by the steep repulsive 
forces between atoms, is characterized in detail, and its consequences
are explored. Beyond the role of attractive forces themselves, 
a key aspect in explaining the differences in the dynamical behavior 
of the two models is the truncation of the interaction potentials 
beyond a cutoff at typical interatomic distance. This leads us to question 
the ability of the jamming scenario to describe the physics of 
glass-forming liquids and polymers.
\end{abstract}


\maketitle

\section{Introduction}

Since van der Waals, the statistical mechanics of liquids has been built on
the idea that the structure of dense liquids 
is predominantly determined by the steep
repulsive forces between molecules and that longer-ranged attractive
interactions merely establish a cohesive background that affects the
thermodynamics, but neither the structure nor the
dynamics~\cite{Longuet-Higgins,Widom,WCA,Hansen-Macdo}. The physical
rationale for this idea is that, in a dense liquid, density fluctuations coming
from the displacement of the atoms or molecules involve the short-ranged
forces. Longer-ranged forces, which are associated with a large number of
particles, average to zero and have thus a negligible contribution to these
fluctuations. This idea has been incorporated into an operational scheme by
Weeks, Chandler and Andersen (WCA)~\cite{WCA} who proposed to split the
intermolecular potential into a short-ranged repulsive component and a
longer-ranged, more slowly varying component, and to treat the latter as a
perturbation. This has provided a powerful tool to study liquids. Especially
for simple atomic liquids, it has led to a very successful description of
the structure, at the level of the static pair density correlations, and of
the thermodynamics~\cite{WCA,WCAreview}.

As already understood in part by van der Waals, and stressed repeatedly in
modern implementations~\cite{WCA,WCAreview,weeksreview,chandlerreview}, 
this picture which neglects
the fluctuations associated with the attractive, and more generally long-ranged, 
interactions becomes less accurate or even breaks down in some cases. For
instance, when the liquid has large-size inhomogeneities, such as an
interface, or when the attractive forces are short-ranged and directional,
the net vector force on a given molecule or on a liquid domain that results
from attraction with the rest of the system no longer vanishes. Then, a
homogeneous, mean-field account of the attractive interactions is no longer
valid. As is also well-known, on the low-density side of the liquid range,
near the critical point or near the gas-liquid spinodal induced by 
attractive forces, density
fluctuations develop on all lengthscales and become large, 
which again makes the mean-field description inappropriate.

If the van der Waals picture and the WCA theory of liquids have been
extensively studied and tested as far as the pair structure and the
thermodynamics are concerned, much less has been undertaken for the
dynamics. Actually, the WCA theory and the associated division of the pair
potential have not been submitted to systematic
investigations~\cite{berne,kushik}. 
We have recently revisited this point in the
context of glass formation and found that a binary Lennard-Jones model and
its WCA reduction to the repulsive components of the pair potentials, when
studied at the same density, show very similar static pair density correlation
functions but widely differing relaxation times as one lowers the
temperature towards the glass phase~\cite{Berthier-Tarjus09}. 
In the present article, we expand on
this first study and present a more exhaustive set of results for both the
structure and the dynamics of the two liquid models. It confirms that a
simple atomic glass-forming liquid in its viscous regime is manifestly an
exception to the van der Waals picture for what concerns its dynamics, but
not its pair structure. Although we cannot unambiguously assign this
observation to a unique physical cause, it seems likely that it indicates
that the viscous liquid is more heterogeneous than the mere consideration of
its pair density correlation function would suggest.

In the previous short account of this work~\cite{Berthier-Tarjus09}, we have
described the effect of the attractive forces, included in the full
Lennard-Jones model but not in its WCA reduction, as ``nonperturbative'' in
the viscous liquid range. We address here more thoroughly this issue. We
try to characterize the \textit{qualitative} difference in dynamical
behavior between the two models, by studying in particular the scaling of
the relaxation time with density and temperature. 
This scaling appears as a ubiquitous
(but approximate) property of glassforming liquids and
polymers~\cite{Alba-et-al,Casalini-Roland,Dreyfus,Fragiadakis-Roland2010}; 
it is obeyed by the Lennard-Jones model but 
is strongly violated in the purely repulsive WCA model.

The presence or absence of density scaling in the dynamics of the binary
Lennard-Jones liquid and its WCA reduction is clearly the consequence of the
presence or absence of an attractive tail in the pair potentials of the two
models. However, one should be cautious about the conclusion to be drawn
from this fact. It is actually not obvious whether this is due to the
attractive nature of the tail \textit{per se} or to the introduction of a
cutoff at a typical interatomic distance in the pair potentials, 
irrespective of
the attractive or repulsive nature of the neglected longer-ranged
interactions. Density scaling is indeed an exact property of purely
repulsive power-law potentials, and it has been recently shown that both the
structure and the dynamics of the binary Lennard-Jones model can be
reproduced by a binary mixture of atoms with purely repulsive power-law
interactions~\cite{Coslovich-Roland,Dyre_PRL2010}. 
On the other hand, as we shall also stress, 
density scaling is strongly violated, at low temperature and
pressure, in systems of repulsive harmonic-like spheres, for which the pair
potentials vanish beyond the typical interatomic 
distance~\cite{berthier-witten1,berthier-witten2}.

Building on these 
observations, we address the relevance of the ``jamming 
scenario''~\cite{Liu-Nagel-et-al} for describing the glass
transition of liquids in their experimentally accessible range of density.
This leads us to question the recently suggested equivalence between jamming
phenomenon and colloidal glass transition on the one hand and glass
transition of supercooled liquids on the other~\cite{Xu-Liu-Nagel}.

In Sec.~\ref{model} we describe the technical detail of our work.
In Sec.~\ref{gr} we contrast the static and dynamic behavior
of the two models under study. We study the scaling with density and
temperature of the dynamics of the models in Sec.~\ref{scaling} and 
we address the role of attractive forces \textit{per se} versus that of introducing a cutoff at a typical interatomic separation in the pair potential in Sec.~\ref{cutoff}. Sec.~\ref{jamming} is devoted to a discussion of the connection between  glass-forming liquids and systems near jamming. Finally, we conclude the paper in Sec.~\ref{conclusion}.

\section{Models, simulation, and phase diagram}
\label{model}

\subsection{Models}

We compare the structure and the dynamics of a standard three-dimensional
model of glass-forming liquid, the Kob-Andersen 80:20 binary Lennard-Jones
mixture~\cite{kob-andersen1994}, and of its reduction to the purely repulsive
part of the pair potentials proposed by WCA.
In what follows, the former will be denoted by ``LJ'' and the latter by
``WCA''. The interatomic pair potential between species $\alpha$ and
$\beta$, with $\alpha, \beta = A,B$, is given in the two systems by
\ba
v_{\alpha \beta}(r)
&=& 4\epsilon_{\alpha \beta}\left[\left( \frac{\sigma_{\alpha
\beta}}{r}\right)^{12}- \left( \frac{\sigma_{\alpha \beta}}{r}\right)^{6} +
C_{\alpha \beta} \right] , \; {\rm for} \; 
r \leq r_{\alpha \beta}^c \nonumber \\
& =& 0, \; { \rm for} \; r \geq r_{\alpha \beta}^c, 
\label{eq_potentials} 
\ea
where $r_{\alpha \beta}^c$ is equal to the position of the minimum of
$v_{\alpha \beta}(r)$ for the WCA model and to a conventional cutoff of $2.5
\sigma_{\alpha \beta}$ (merely introduced for practical reasons with no
impact on the physical quantities) for the standard LJ model; $C_{\alpha
\beta}$ is a constant that is fixed such that $v_{\alpha \beta}(r_{\alpha
\beta}^c) =0$. The difference between the LJ and WCA potentials is a purely
attractive contribution, dominated  by the $r^{-6}$ contribution at large
distance.

We have performed Molecular Dynamics simulations in the $NVE$ ensemble
(after equilibration at a chosen temperature) with $N=900-1300$ 
particles (depending on density) and we
have studied a broad range of density $\rho$ from $1.1$ to $1.8$.
Lengths, temperatures and times are given in units of $\sigma_{AA}$,
$\epsilon_{AA}/k_B$, and $(m \sigma_{AA}^2/48 \epsilon_{AA})^{1/2}$
respectively. It should be noted that the WCA truncated potential
is continuous at the cutoff, but the resulting forces are not, which 
leads to a slow drift of the total energy during very long 
simulation runs in the microcanonical ensemble. To cure this problem 
without introducing random collisions or a thermostat, 
we periodically rescale velocities with a very low frequency 
to maintain the total energy constant~\cite{carre}. 

To check the robustness of our results, we have also 
considered the two-dimensional version of the LJ and the WCA models, using 
a 65:35 mixture, because it is known to be less prone
to crystallization in two dimensions than the 80:20 mixture~\cite{2d}. 

\subsection{Thermodynamics and phase diagram}

\begin{figure}
\psfig{file=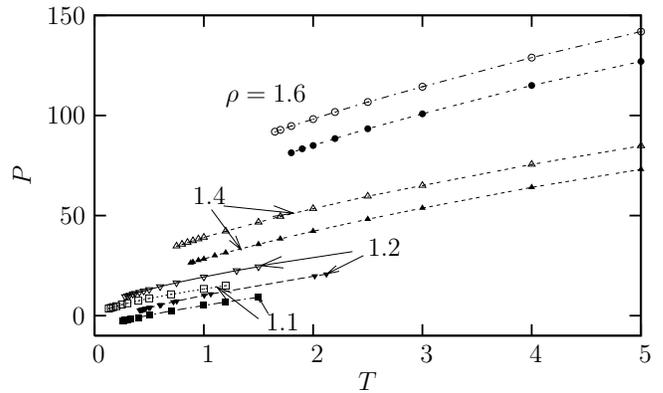,width=8.5cm}
\caption{\label{figure_pressure} The $P(T)$ equation of
state for the LJ (filled symbols) and WCA (open sybols)
models at different densities merely differ by a 
constant.}
\end{figure}

In the WCA theory, the repulsive truncated model is taken as a reference for
the full LJ model, from which one can determine the structure of the liquid,
whereas the attractive component is treated as a perturbation that allows
one to compute thermodynamic quantities. Within this description, the
structure and the dynamics of the LJ and WCA models should be compared at
the same ($\rho, T$) state point. Their pressure then differs, with the
attractive interaction roughly providing a negative background term. This is
illustrated in Fig.~\ref{figure_pressure} where we plot the simulation
results for the equation of state, pressure versus temperature $P(T)$, 
of the two
models at four densities from $\rho=1.1$ to $\rho=1.6$: the pressure of the
WCA model is roughly shifted up by a $\rho$-dependent constant from that of
the full LJ model. As can be seen, the pressure of the full LJ mixture becomes
negative at the lowest temperatures for $\rho=1.1$, which is the sign that
the liquid is in a metastable state in the two-phase region inside the
gas-liquid coexistence curve. The purely repulsive WCA model on the other
hand has always a positive pressure and no coexistence region.

\begin{figure}
\psfig{file=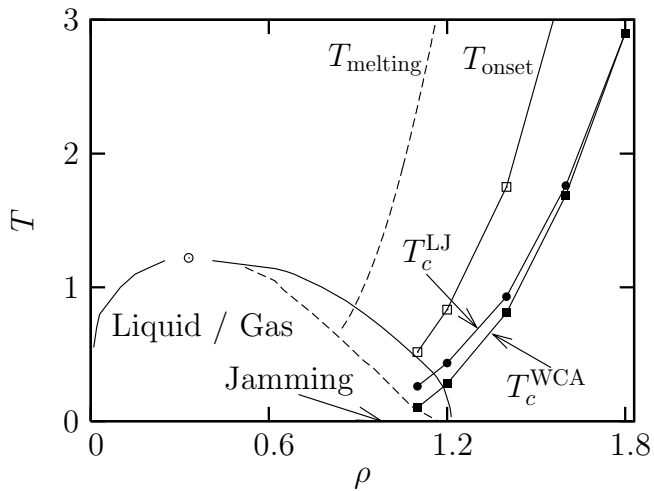,width=8.5cm}
\caption{\label{figure_phase_diagramLJ} Phase diagram of the system with:
the melting line of the monodisperse LJ system~\cite{melting}, the 
coexistence curve~\cite{binodal}, the spinodal line~\cite{spinodal} 
and the onset temperatures $T_{\rm onset}$ discussed in 
Sec.~\ref{scaling} of the 
binary LJ mixture, 
the $T \to 0$ jamming limit of the WCA model; 
the mode-coupling transition line $T_c(\rho)$ as fitted from the 
numerical data~\cite{Berthier-TarjusMCT}.} 
\end{figure}

To help characterizing the different state points that will be considered,
we give in Fig.~\ref{figure_phase_diagramLJ} the phase diagram in the
($\rho,T$) plane of the binary LJ model. For illustrating the relevant range
for the dynamics, we also plot on this diagram the estimated $T_c$ of the
mode-coupling theory as obtained from a fit to the simulation
data~\cite{Berthier-TarjusMCT}. For the purpose of this work,
the merit of such a fit is immaterial, the $T_c$ line being 
a convenient indicator of a temperature scale where dynamics
has slowed down by about 4 decades as compared to the high-temperature fluid. 
The mode-coupling line ends up for $\rho=1.1$ in the metastable
liquid, as mentioned above, whereas densities $\rho=1.6$ and
$\rho=1.8$ are clearly outside the conventional liquid range (one is always
much above the critical temperature). These densities 
actually correspond to conditions that, although easily 
investigated in simulations of simple
models, are not experimentally relevant, even in high-pressure experiments
on glass-forming liquids (see also Ref.~\cite{Voigtmann}). The
density $\rho=1.2$ is the standard condition at which the slowdown of
relaxation of the binary Lennard-Jones mixture is usually 
considered~\cite{kob-andersen1994}. It corresponds to typical
liquid (and viscous liquid) states. Most of our simulations are performed
in the viscous regime in the region 
delimited by the onset temperature $T_{\rm onset}$ and the mode-coupling
lines in Fig.~\ref{figure_phase_diagramLJ}.

As for the WCA model, because of the absence of attractive interactions, it
does not have a gas-liquid transition and could in principle be studied at
all densities down to zero temperature. In practice however, we found 
that the system crystallizes in simulations when cooled down to 
small but nonzero $T$ at densities below $\rho \simeq 1.1$, and 
this is the reason which prevents us from 
extending the mode-coupling line down to $T=0$.  
On the other hand, directly at $T=0$, the WCA model could in principle be
subjected to various ``jamming'' protocols with density as the control
variable~\cite{OHern-Liu-Nagel}. We shall come back to these aspects in the
following.

\section{Contrasting the behavior of the pair structure and of the dynamics}
\label{gr}

\subsection{Pair correlation functions}

We first consider the structure of the two liquid models, as characterized
by the static two-body density correlations. We focus here on the pair
correlation functions $g_{\alpha \beta}(r)$, the static structure factors
$S_{\alpha \beta}(q)$ having already been displayed
elsewhere~\cite{Berthier-TarjusMCT}.

\begin{figure}
\psfig{file=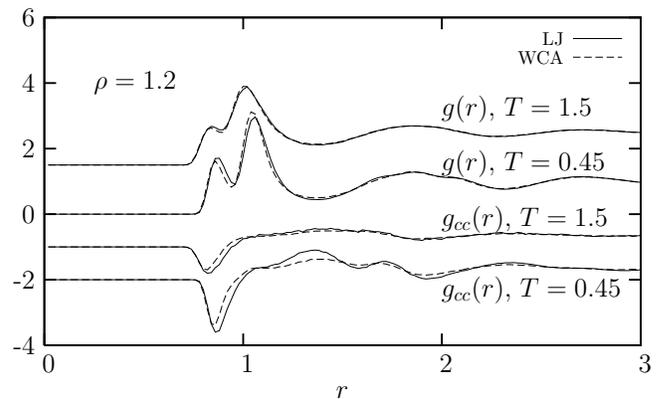,width=8.5cm}
\caption{\label{figure_density_concentration} Pair correlation functions of
the (total) density fluctuations $g(r)$ and of the concentration
fluctuations $g_{cc}(r)$ for the LJ and WCA liquids at high and low
temperature for the canonical liquid density $\rho=1.2$. The different curves 
are vertically shifted, for clarity.}
\end{figure}

\begin{figure*}
\psfig{file=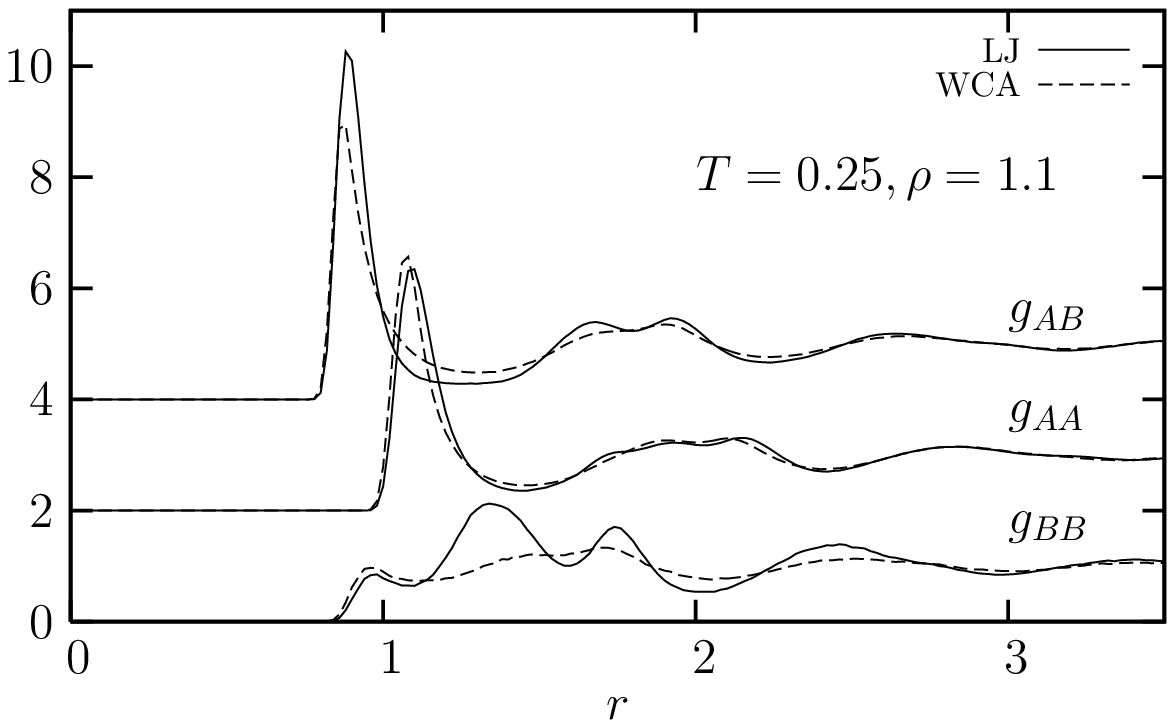,width=8.5cm}
\psfig{file=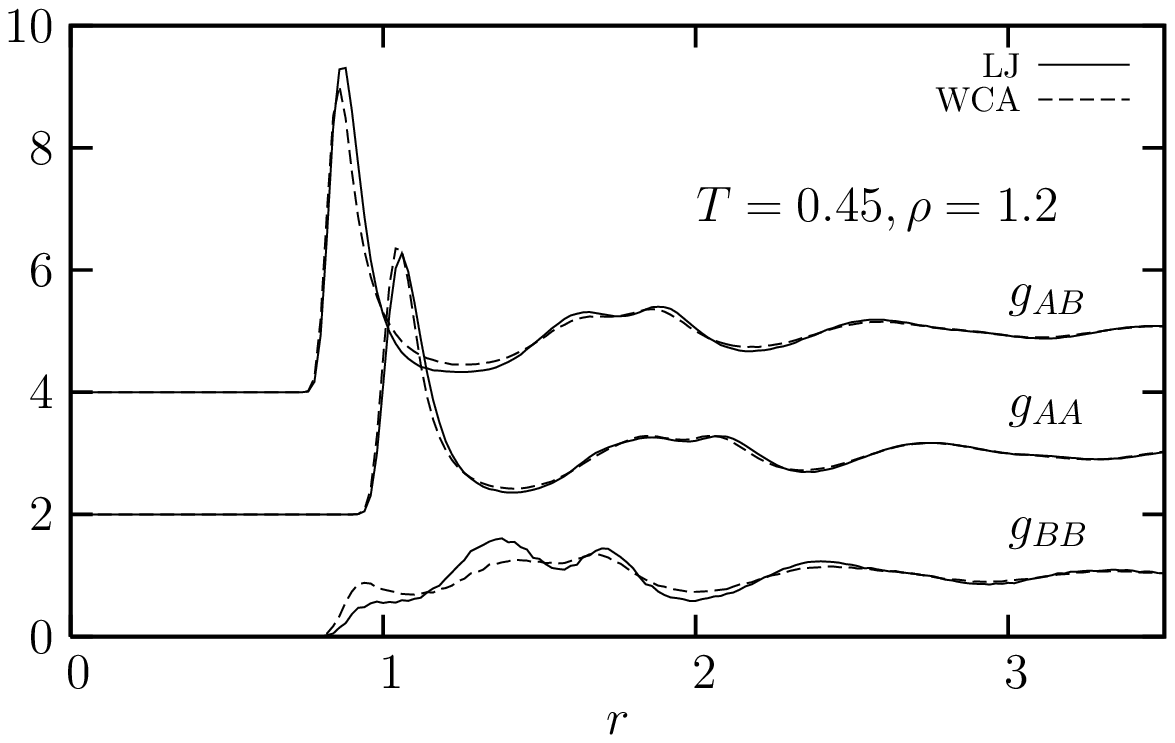,width=8.5cm}
\psfig{file=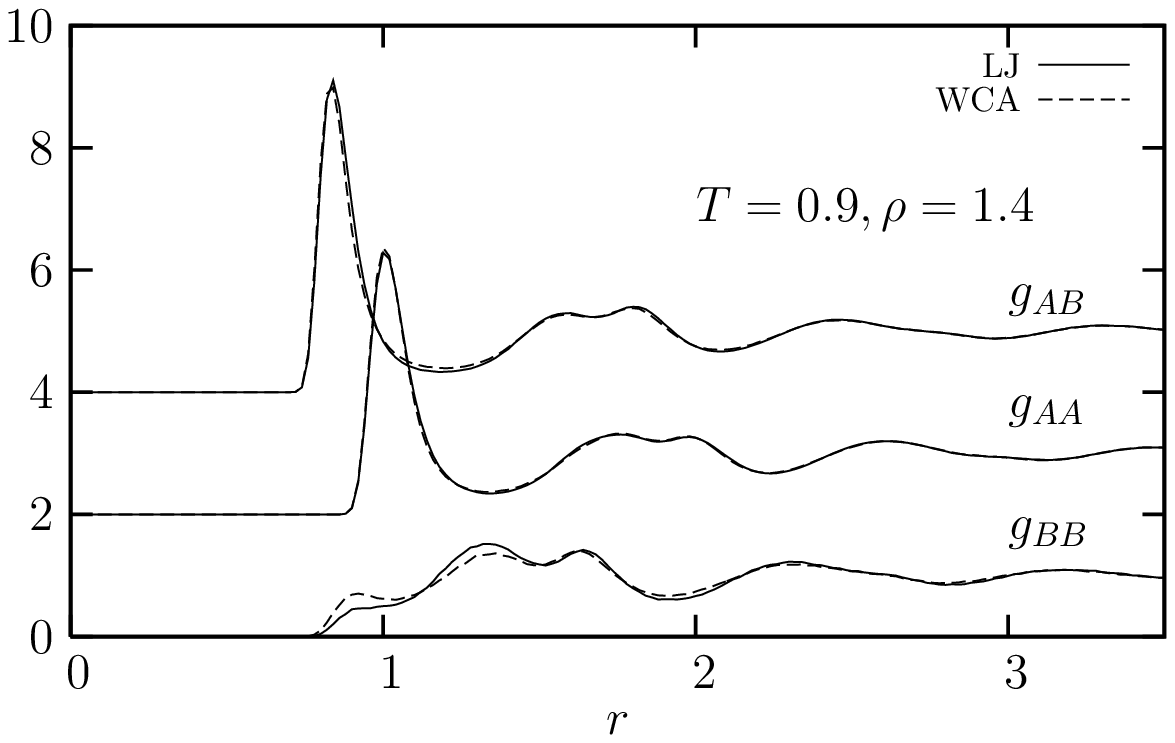,width=8.5cm}
\psfig{file=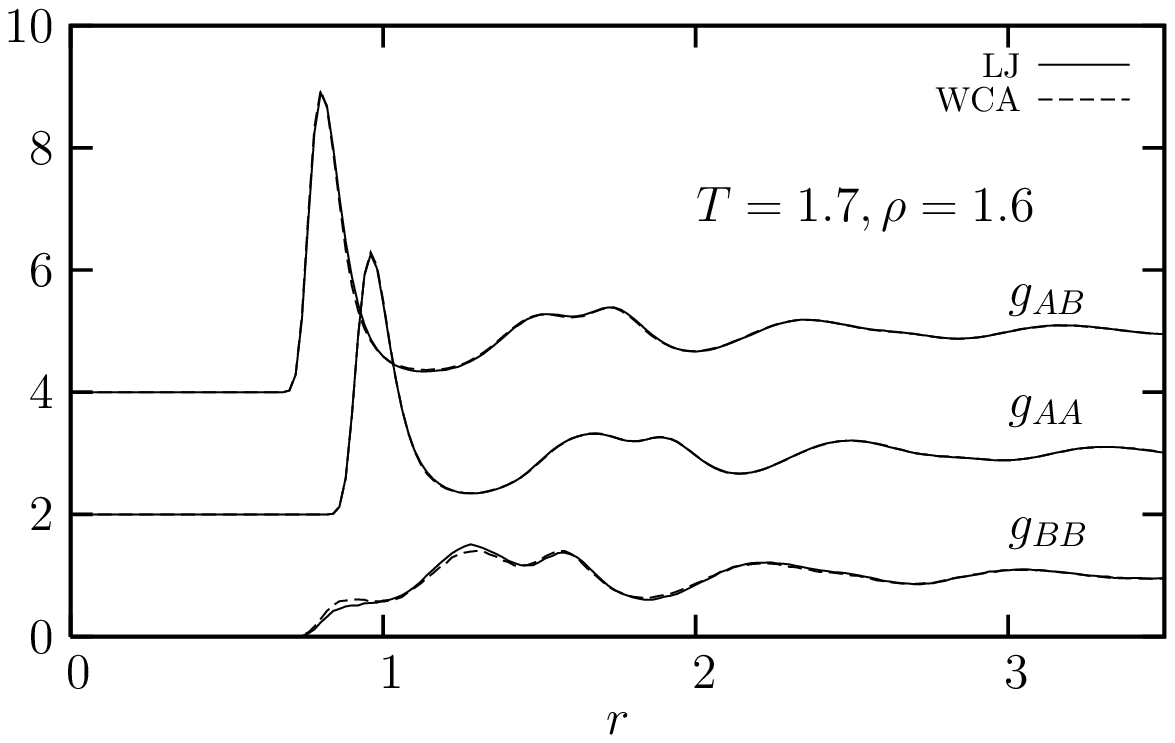,width=8.5cm}
\caption{\label{figure_partials} Evolution with density of the partial pair
correlation functions of the LJ and WCA liquids at low temperature
and various densities. The different curves 
are vertically shifted, for clarity.}
\end{figure*}

In Ref.~\cite{Berthier-Tarjus09}, we have shown that the pair correlation
function of the (total) density fluctuations, 
\begin{equation}
\label{eq_density_g(r)} 
g(r) \equiv x_A^2 \,
g_{AA}(r)+2\, x_A x_B \, g_{AB}(r)+x_B^2 \, g_{BB}(r), 
\end{equation} 
where
$x_{\alpha}$ denotes the concentration of species $\alpha$, is extremely
similar for the LJ and the WCA models and that its shape varies only weakly
with temperature. This is illustrated in
Fig.~\ref{figure_density_concentration} for the canonical density
$\rho=1.2$. Weeks, Chandler and Andersen~\cite{WCA} pointed out that, in the
case of a binary mixture, the coincidence between the structure of the two
models should not be as good for the pair correlations involving the
fluctuations of concentration ($\delta c=x_B \delta\rho_A-x_A
\delta\rho_B$), as those are less constrained at high density than the
(total) density fluctuations ($\delta \rho=x_A \delta\rho_A+x_B
\delta\rho_B$). We have checked this by plotting the pair correlation
function of the concentration fluctuations, $g_{cc}(r)$, defined as
\begin{equation} 
\label{eq_concentration_g(r)} 
g_{cc}(r) \equiv x_B^2 \,
g_{AA}(r)-2\, x_A x_B \, g_{AB}(r)+x_A^2 \, g_{BB}(r). 
\end{equation} 
The
results are also 
shown in Fig.~\ref{figure_density_concentration}. As anticipated,
the difference between the LJ and WCA models is still weak but more
pronounced for $g_{cc}(r)$ than for $g(r)$.

Finally, we display in Fig.~\ref{figure_partials} the evolution with density of the
partial pair correlation functions $g_{\alpha \beta}(r)$ for the two liquid
models. As there is less difference between the two
models at high temperature, we concentrate on the low temperatures, thus
emphasizing those state points where differences are more pronounced.
At high
density, all curves nearly superimpose, 
but one can see notable differences between
the two models at $\rho=1.1$, especially for the correlation functions
involving the minority species $B$. At such a density, attractive
interactions start playing a nonnegligible role as the full LJ system enters
the metastable liquid region inside the coexistence curve and approaches the
spinodal where density fluctuations grow very large. 

It is perharps  not
surprising that $BB$ correlations are more affected by removing the
attractive part of the potential, as the interaction parameters of the 
model were specifically chosen to favor attraction between $AB$ 
particles, such that $B$ particles can efficiently 
frustrate the crystallization
of the majority $A$ component. By removing the attractive interactions, we see that the first 
peak of $g_{BB}(r)$ is more pronounced for the WCA model, 
while the first peak in $g_{AB}(r)$ is less pronounced, 
showing that the effect engineered by Kob and Andersen 
is less efficient. This 
also explains why the present WCA mixture crystallizes more easily at low 
density than its LJ counterpart. Note finally that $g_{BB}(r)$
dominates the behavior of $g_{cc}(r)$, see Eq.~(\ref{eq_concentration_g(r)}), 
which explains why $g_{cc}(r)$ is a more sensitive probe of the 
structural differences of the two models than $g(r)$, which 
is dominated by $g_{AA}(r)$.  

Although we discussed in detail these small differences, 
our results on the whole 
provide one more confirmation of the validity of the WCA theory
for the equilibrium pair structure of dense liquids. The rather small
differences observed between the LJ and WCA models could 
probably be captured by a perturbative treatment of
the attractive tail, for instance  along the lines suggested
in Ref.~\cite{Chandler-OCT}.

\subsection{Dynamic correlations and timescales}

\begin{figure}
\psfig{file=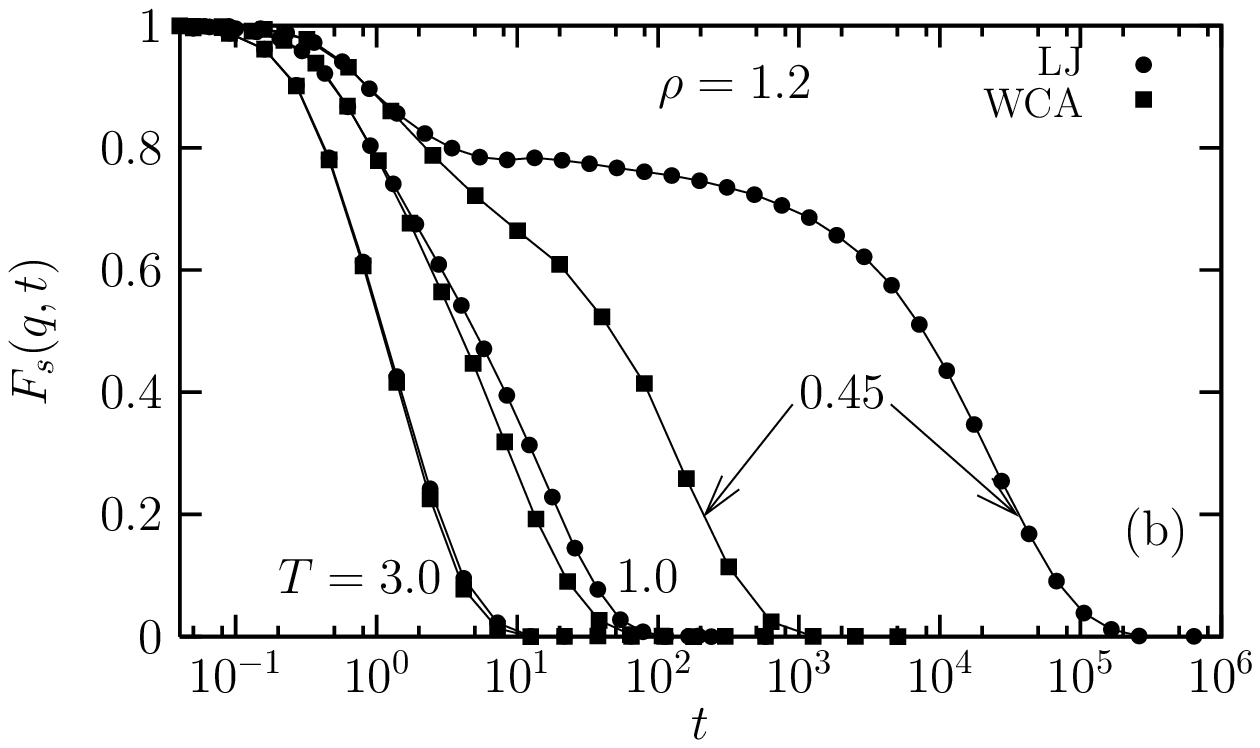,width=8.5cm}
\psfig{file=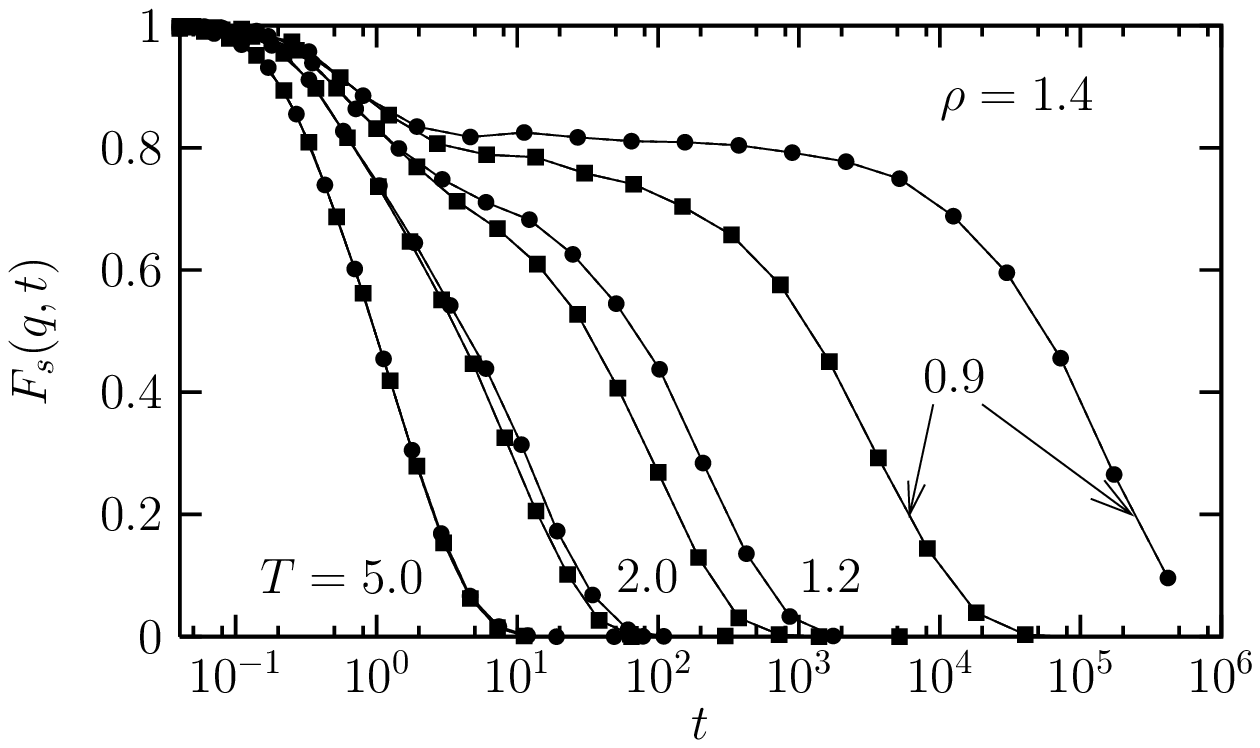,width=8.5cm}
\psfig{file=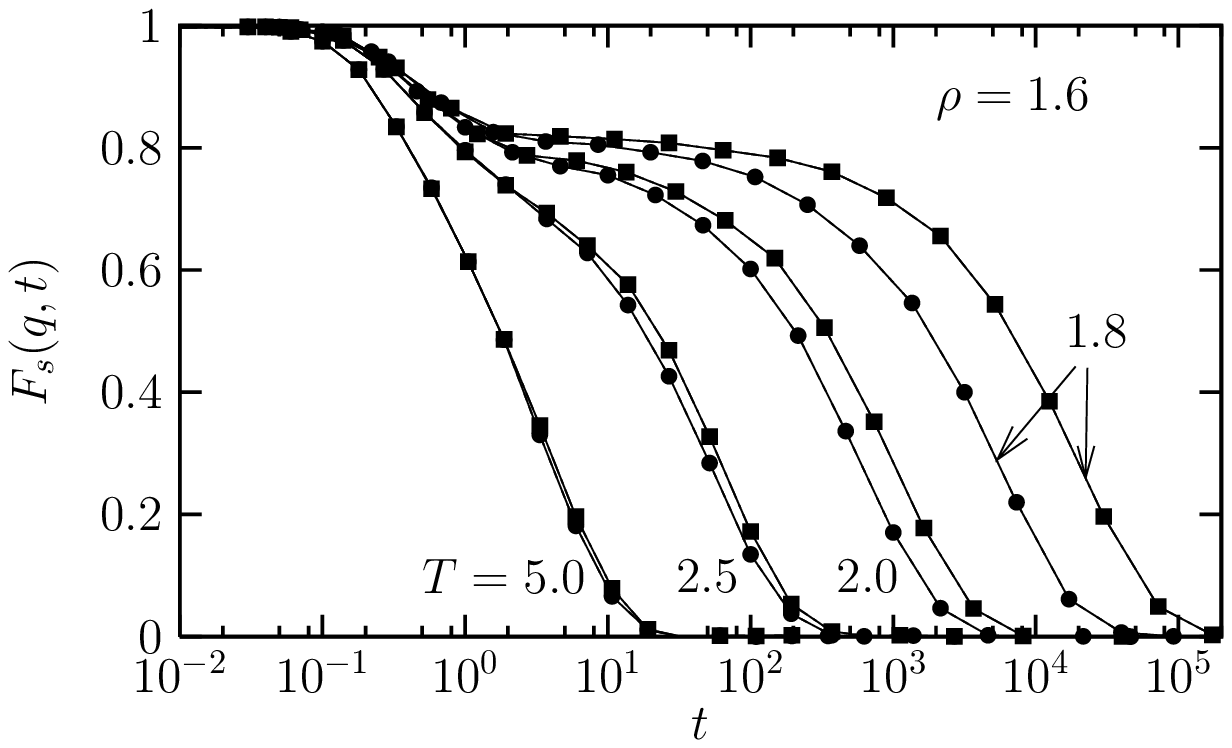,width=8.5cm}
\caption{\label{figure_self_intermediate_scattering}
Comparison of the time dependence of the self-intermediate 
scattering function $F_s(q,t)$ for
$q \sigma_{AA}\simeq 7.2$ of the LJ and WCA models at the same temperatures,
shown for different densities. }
\end{figure}

We now turn to the results concerning the dynamics of the two liquid models.
As already shown in our previous publication~\cite{Berthier-Tarjus09}, the
idea that the dynamical behavior of a liquid is essentially insensitive to
longer-ranged attractive interactions breaks down in an increasingly
spectacular manner as temperature decreases. The difference between the
liquid in the presence and in the absence of the attractive tail is small
but already noticeable at high temperature in the ``normal'' range (see also
Refs.~\cite{kushik,berne})  and it rapidly builds up in the low-temperature regime relevant for the 
glass transition phenomenon~\cite{Berthier-Tarjus09}. Here,
we analyze in more detail the influence of the density.

To analyze the dynamics of the models, 
we have mostly studied the time dependence of the 
self-intermediate scattering functions defined by 
\begin{equation}
\label{eq_self_intermediate} 
F_s(q,t)=\frac{1}{N} \left\langle \sum_{j=1}^{N} e^{i
\mathbf{q}.(\mathbf{r}_j(t)-\mathbf{r}_j(0))} \right \rangle, 
\end{equation} 
with $q \sigma_{AA}\simeq 7.2$, which we kept constant 
for all state points. This corresponds roughly to studying 
single particle displacements over a length scale comparable to 
the inter-particle distance.
Representative data points are shown in 
Fig.~\ref{figure_self_intermediate_scattering} for three densities
and a broad range of temperatures. 
These figures strikingly illustrates how the small differences
in the relaxation dynamics of the high-temperature 
liquids become a dramatic effect at low temperatures. For $\rho=1.2$ and
$T=0.45$ for instance, the LJ system hardly relaxes in the timescale
of the simulation while this temperature corresponds 
to a modestly supercooled state. When $\rho$ increases, these 
differences decrease, but note that even for $\rho=1.6$ and $T=1.8$ 
(a temperature which is 50 \% larger than the critical point, see 
Fig.~\ref{figure_phase_diagramLJ}) the time correlation
functions still differ by a large factor. 

\begin{figure}
\psfig{file=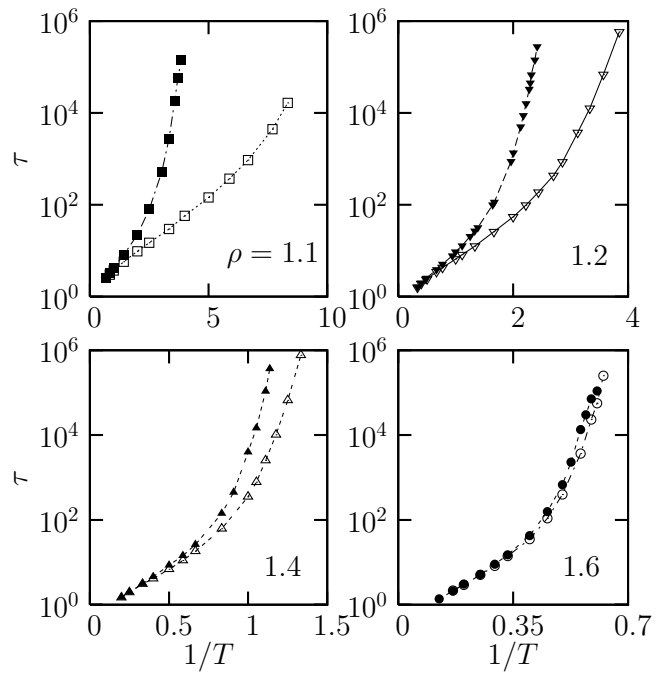,width=8.5cm}
\caption{\label{figure_relaxation_time} Arrhenius plot of the relaxation
time of the LJ (filled symbols) and WCA (open symbols) models for several
densities. Note that the temperature range is different for each panel.}
\end{figure}

From the time decay of the self-intermediate scattering function, 
we extract a relaxation time, $\tau$, defined in practice 
as $F_s(q,\tau) = \exp(-1)$. We show in
Fig.~\ref{figure_relaxation_time} an Arrhenius plot of this relaxation time
for several densities. The figure clearly illustrates that the
dramatic difference between the two models that is seen at liquid densities
(with already a difference of 3 orders of magnitude at the lowest
temperature at which we can equilibrate the full binary LJ model) decreases
slowly with density. In such a logarithmic representation covering 
many decades, the presence or absence of
the attractive tail of the potentials appears irrelevant 
at very high densities only. However, as stressed above, 
such densities are not realistic for actual glass-forming liquids.

\subsection{Two-dimensional case and other examples}

\begin{figure}
\psfig{file=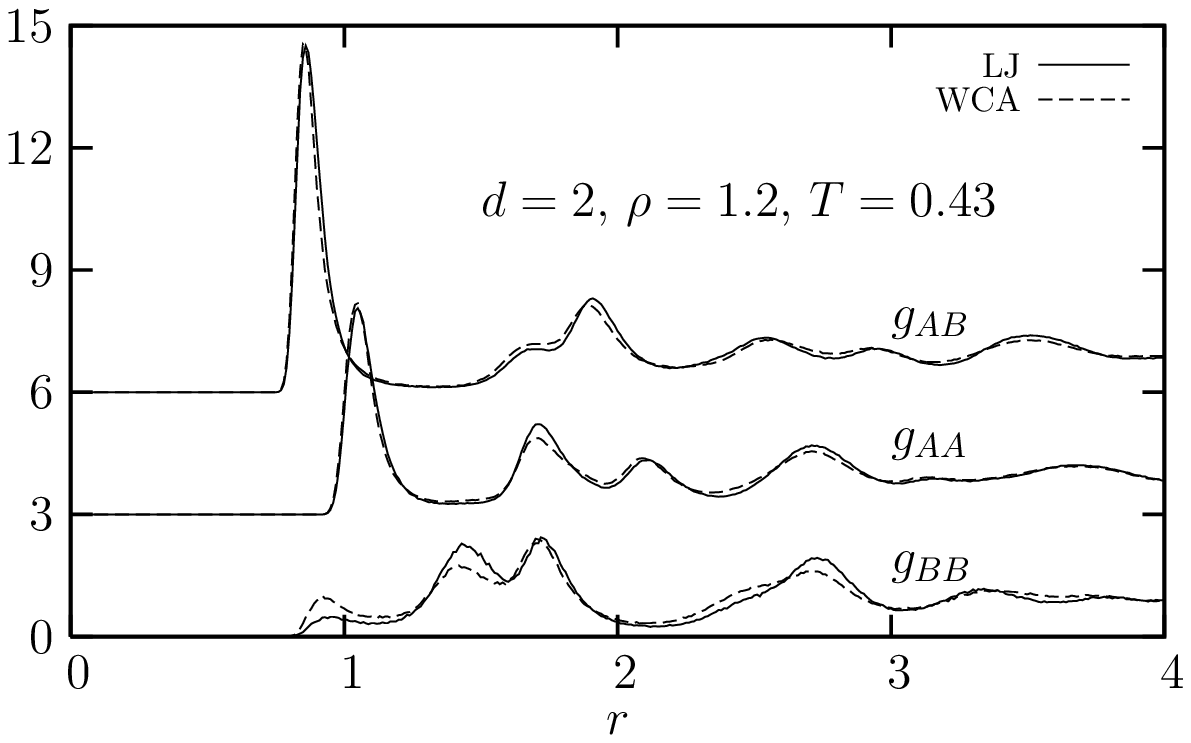,width=8.5cm}
\psfig{file=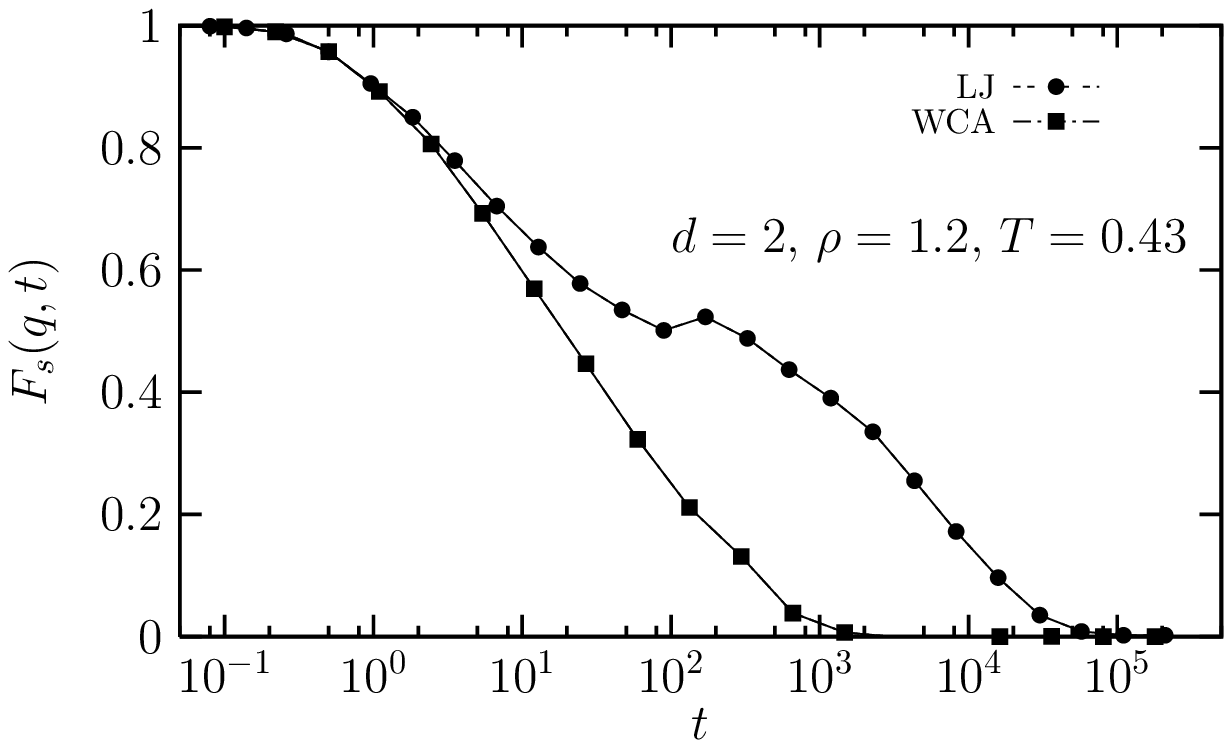,width=8.5cm}
\caption{\label{figure_2D} Two-dimensional 65:35 Kob-Andersen binary LJ mixture
and its WCA reduction: Top: Partial pair correlation functions $g_{\alpha
\beta}(r)$ for a typical liquid density $\rho=1.2$ at low temperature.
Bottom: Time dependence of the self intermediate scattering functions under
the same conditions.}
\end{figure}

We note in passing that the pattern that we have found here, namely that the
WCA reduction of the binary LJ mixture to a truncated purely repulsive
systems has virtually no effect on the pair correlations of the total
density but strongly affects the dynamics, is not unique. We have also
obtained the same effect for instance in a two-dimensional version of 
the binary 65:35 LJ mixture model, as illustrated in 
Fig.~\ref{figure_2D}.

We point out that, although not appreciated before,
the conclusion that the WCA truncation of Lennard-Jones forces 
has a dramatic impact on the dynamics 
can in fact be drawn by comparing the results
already published in the literature for different Lennard-
Jones mixtures~\cite{kob-andersen1994,lacevic} 
and for their WCA truncations~\cite{maibaum,castillo}.

Additionnally, since their initial publication~\cite{Berthier-Tarjus09},
our results for the three-dimensional Kob-Andersen 80:20 mixture
have now been confirmed by independent 
studies~\cite{Dyre_PRL2010,Coslovich-Roland}.

\subsection{To be, or not to be (perturbative)}

Before closing this section, we would like to discuss how to best
characterize the difference in the dynamical behavior of the LJ and WCA
models at liquid densities and what to make of it. In our earlier
publication~\cite{Berthier-Tarjus09}, we have described the effect of the
attractive tail of the pair potentials on the dynamics as
``nonperturbative''. At a quantitative level, the gap between the
characteristic time scales of the two models indeed becomes enormous and
widens as temperature decreases. By itself, this observation undoubtedly
invalids any claim that attractive tails can be neglected when describing
the dynamics (the relevance, or not, of the attractive nature of the tail
will be discussed in section~\ref{cutoff}), but it is not enough to justify the
qualifier ``nonperturbative''. In principle, this quantitative difference,
no matter how large, could be scaled out through the introduction of
parameters which themselves could be computed via a perturbative treatment
of the longer-ranged interactions. This would be true for instance if the
dynamics could be predicted on the sole basis of the knowledge of the pair
structure, with minor effects in the latter being strongly amplified in the
former. However, we show elsewhere~\cite{Berthier-Tarjus2} that such
approaches are unable to account for the diverging behavior of the LJ and
WCA liquids in the viscous regime as temperature is lowered. In addition, we
have found evidence that the difference in the dynamics of the two systems
is not only quantitative but also qualitative. This is the point we
now discuss in more detail.

\section{On the density scaling for the relaxation time}
\label{scaling}

\subsection{Density scaling in supercooled liquids and polymers}

The simplest way for trying to scale out the difference in the
$T$-dependences of the LJ and WCA liquids is to renormalize the temperature
by one density-dependent parameter chosen to make the curves of the two
models coincide at high temperature, where a perturbative treatment of the
attractive tails more likely to apply.
 
The most convenient way to proceed
is to fit the high-$T$ data to an Arrhenius form,
\be
\tau(\rho,T) \approx \tau_{\infty}(\rho) 
\exp[ E_{\infty}(\rho)/T], 
\ee
which indeed provides
a good empirical description, irrespective of the physical 
meaning one is willing to put in its use. This procedure suggests 
to rescale $T$ for the whole range under study by the
extracted energy parameter $E_{\infty}(\rho)$. When applied to glass-forming
liquids and polymers~\cite{Alba-et-al}, this procedure has been shown to
provide a very good collapse of the relaxation and viscosity data at
different densities, according to 
\begin{equation}
\label{eq_density_scaling1} 
\frac{\tau(\rho,T)}{\tau_{\infty}}=
{\mathcal F}_1 \left[ \frac{E_{\infty}(\rho)}{T} \right]. 
\end{equation} 
With specific functional
forms for the density-dependent scaling parameter (see also below), this
relation has been successfully applied to an impressive variety of
glass-forming liquids and
polymers~\cite{Alba-et-al,Casalini-Roland,Dreyfus,Fragiadakis-Roland2010}. 
Thus, there is 
enough empirical evidence to take this scaling property 
as a genuine characteristic of glass formation 
in these systems. 

It must be noted that Eq.~(\ref{eq_density_scaling1}) is 
far from being trivial, as it connects the dynamics 
in the high-temperature liquids to the behaviour found in the viscous,
low temperature regime. More precisely, Eq.~(\ref{eq_density_scaling1})
shows that two physical quantities are in fact proportional: 
the high-temperature activation energy, $E_\infty$, and the 
``onset temperature'' for slow dynamics, $T_{\rm onset}$, which could be 
operationally defined at the point where ${\cal F}_1(x)$ departs from 
its high-$T$ (Arrhenius) behavior:
\be
T_{\rm onset} (\rho) \propto E_\infty(\rho).
\label{onset}
\ee
Physically, $T_{\rm onset}$ marks the 
temperature below which the liquid starts behaving in a more
collective and heterogeneous 
manner~\cite{Sastry-Deben-Stillinger,KIvelson-Kivelson-Tarjus95,Garrahan-Chandler,moreonset}.

As expected, and already displayed in Ref.~\cite{Berthier-Tarjus09}, the
data for the full LJ binary mixture obey very well the density scaling of
Eq.~(\ref{eq_density_scaling1}), see Fig.~2a of 
Ref.~\cite{Berthier-Tarjus09} (a
small deviation from the data collapse can be observed at the lowest density
$\rho=1.1$ corresponding to the metastable liquid inside the gas-liquid
coexistence curve). On the other hand, the scaling relation is strongly
violated for the WCA liquid: see Fig.~2b of Ref.~\cite{Berthier-Tarjus09}. 

One important consequence is that the isochoric fragility, which
characterizes the steepness of the $T$-driven slowdown of relaxation and can
be quantified by the derivative $\partial
\log[\tau(\rho,T)/\tau_{\infty}]/\partial \log T$ taken at constant $\rho$
and evaluated at the glass transition (or for any given value of
$\tau/\tau_{\infty}$), is essentially independent of density for the LJ
model, as also found in real glass-forming liquids and
polymers~\cite{Alba-et-al}, but strongly depends on 
density for the WCA model.
This presence or absence of (even approximate) density scaling represents,
we claim, a \textit{qualitative} difference between the two liquid models.

\subsection{Empirical data collapse}

If one wishes to also collapse the relaxation data at all densities for the
WCA model, one must introduce at least one 
additional, $\rho$-dependent parameter.
This procedure may have only little physical significance \textit{per se},
but it allows one to quantify and discuss more precisely 
the deviation from density scaling, which is obeyed by the LJ model only.

\begin{figure} 
\psfig{file=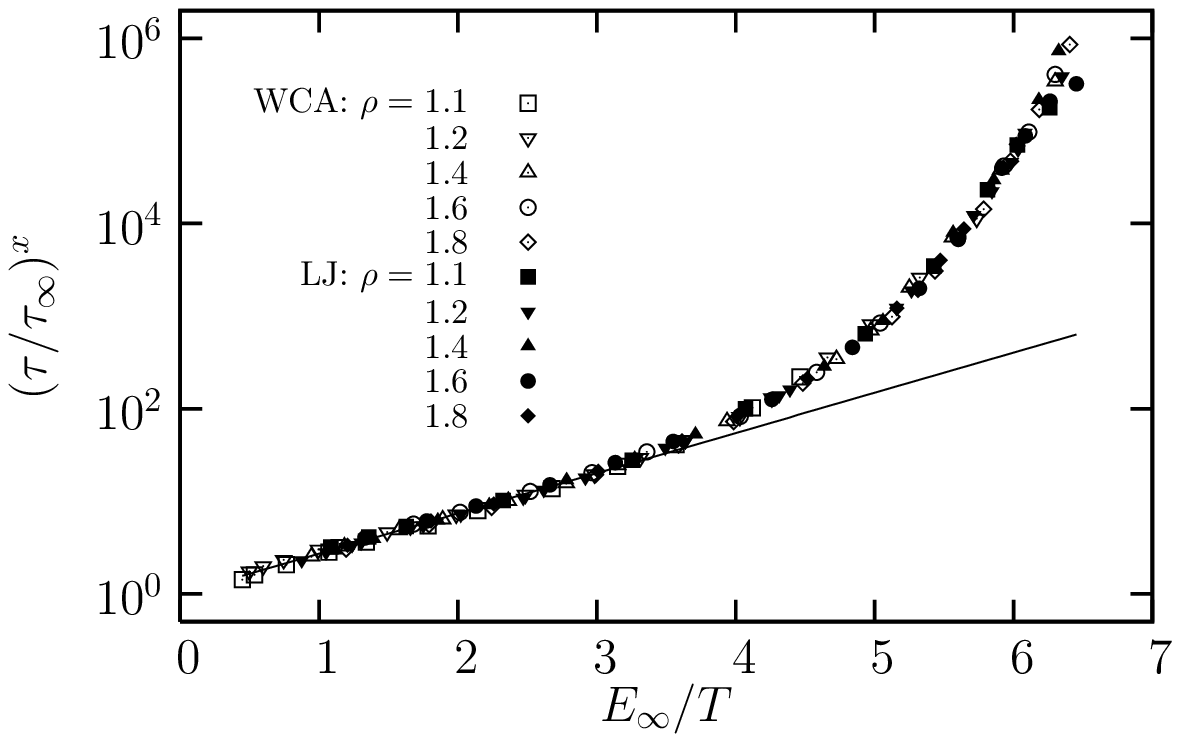,width=8.5cm}
\psfig{file=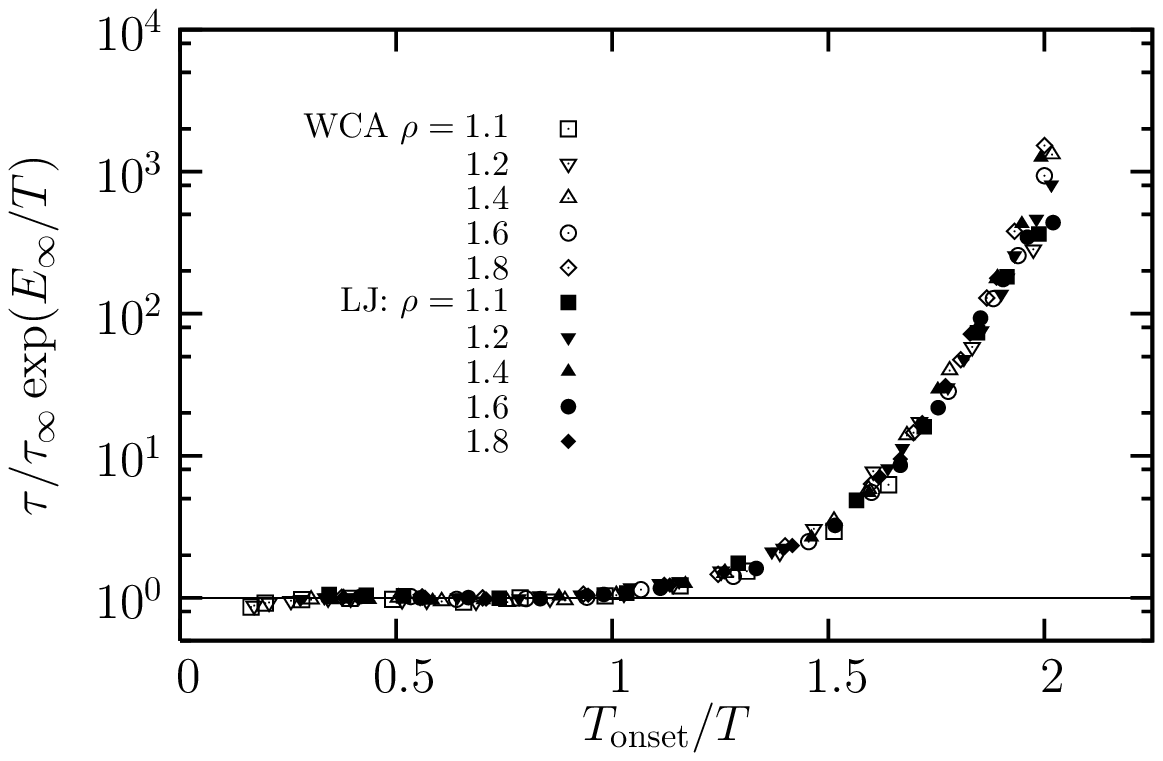,width=8.5cm} 
\caption{\label{figure_collapse} Empirical  
collapse of the relaxation-time data for the LJ and WCA models at densities
from $\rho=1.1$ to $\rho=1.8$. 
Top: Using Eq.~(\ref{eq_density_scaling3}) with parameters listed
in Table~\ref{table}.
Bottom: Using Eq.~(\ref{eq_density_scaling2}) with parameters 
listed in Table~\ref{table}.
The absolute scale of $y(\rho)$ is chosen so
that the deviation from the high-$T$ behavior takes place around
$T_{\rm onset} = y E_{\infty} \approx 1$. }
\end{figure}

We have found that the two functional forms, 
\begin{equation}
\label{eq_density_scaling2}
\frac{\tau(\rho,T)}{ \tau_{\infty} \exp(E_{\infty}(\rho)/T ) } 
=
{\mathcal G}_1 \left[ y(\rho) \frac{E_{\infty}(\rho)}{T} \right] 
\end{equation} 
and
\begin{equation} 
\label{eq_density_scaling3}
\left( \frac{\tau(\rho,T)}{\tau_{\infty}} \right)^{x(\rho)} 
= {\mathcal G}_2 \left[ \frac{E_{\infty}(\rho)}{T} \right], 
\end{equation} 
both lead to
an excellent collapse on a master-curve of the LJ and WCA data, as seen in
Fig.~\ref{figure_collapse}. It is straightforward to show that
the density scaling of Eq.~(\ref{eq_density_scaling1}) is recovered
when either $x(\rho)$ or $y(\rho)$ is independent of density.

We arrived at the formula in 
Eqs.~(\ref{eq_density_scaling2}, \ref{eq_density_scaling3})
by using the possibility that Eq.~(\ref{onset}) is not valid 
and that one needs two distinct energy scales to 
describe the behaviour of the WCA model. This additional freedom can 
be incorporated in a variety of ways, and we present two 
of them in Fig.~\ref{figure_collapse}. 
In a recent work~\cite{Elmatad-Chandler}, our 
published data~\cite{Berthier-Tarjus09} for the LJ and WCA 
models have been re-analyzed and 
collapsed for all densities
by means of a specific functional form which similarly makes use 
of two independent energy scales; however, this collapse 
only holds in the viscous regime, while we also consider the 
high-temperature regime in Fig.~\ref{figure_collapse}.

\begin{table}
\begin{tabular}{|| l || l | l | l | l || l | l | l | l ||}
     & WCA & & & & LJ &  &  & \\
\hline
$\rho$  & $T_{\rm onset}$ &  
$E_\infty$ & $y$ & $x$ & $T_{\rm onset}$ 
& $E_\infty$  & $y$ & $x$ \\
\hline 
1.1  &  0.197   & 0.885 & 4.49 &  0.586   & 0.517   & 1.6  &  3.09 & 1  \\
1.2  &  0.513   & 2.    & 3.89 &  0.685   & 0.833   & 2.55 &  3.06 & 1    \\
1.4  &  1.51    & 5.25  & 3.47 &  0.897   & 1.75   & 5.6  &  3.20 & 1    \\
1.6  &  3.2     & 10.6  & 3.31 &  0.958   & 3.33   & 10.9 &  3.27 & 1    \\
1.8  &  5.6     & 18.   & 3.21 &  1.0     & 5.67     & 18   &  3.17 & 1    \\
\hline
\end{tabular}
\caption{\label{table} The parameters for the empirical data collapse. The
parameter $y$ is equal to the ratio between $E_\infty$ and
$T_{\rm onset}$. It fluctuates around $\approx 3.2$ for LJ 
at all densities, and decreases strongly with $\rho$ for the WCA.}
\end{table}

As expected, we find that for the LJ model, the two
additional parameters $y(\rho)$ and $x(\rho)$ are indeed 
essentially independent of
density. While $x=1$ is used in Fig.~\ref{figure_collapse},
the $y$ values are given in Table~\ref{table}, with small 
variations from a constant that have no systematic dependence
on density within statistical accuracy.
On the contrary, to collapse also the data for the WCA model, 
significant deviations from constant behavior (with systematic 
trends as a function of density) are needed
for both $x(\rho)$  and
$y(\rho)$, see Table~\ref{table}. 

The presence or absence of density scaling for LJ and WCA 
models is a qualitative effect, which we have empirically quantified 
by introducing the parameter $x$ or $y$
in Eqs.~(\ref{eq_density_scaling2}, \ref{eq_density_scaling3}).
As pointed out in Ref.~\cite{Elmatad-Chandler}, the differences
found in the parameters obtained for the two models are modest 
($x$ or $y$ vary by about 50 \%) compared to the observed differences 
in the relaxation times themselves; but this is of
course expected for parameters characterizing the 
temperature dependence of the
\textit{logarithm} of the relaxation times, and this does not contradict our
conclusion that the presence or absence of density scaling represents a
qualitative feature of the slowing down of a glass-former.

\begin{figure}
\psfig{file=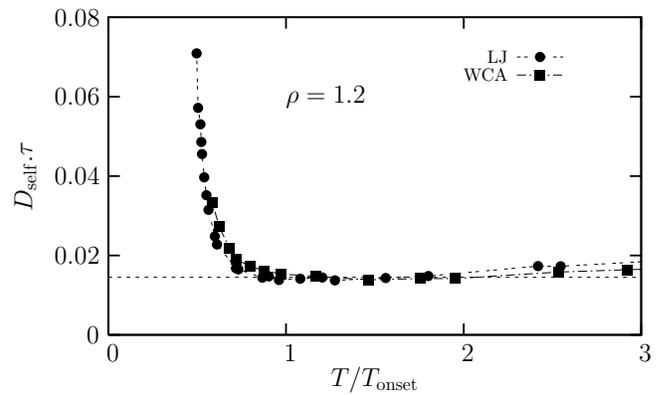,width=8.5cm}
\caption{\label{figure_decoupling} Decoupling between the $T$ dependences of
the self diffusion constant $D_{\rm self}$ and of the relaxation time $\tau$
for $\rho=1.2$ as a function of the temperature $T$ rescaled 
by the independently measured $T_{\rm onset}$. 
For both models, the product $D_{\rm self}\, \tau$ begins to increase below
$T \approx T_{\rm onset}$, which differ by a factor 
1.6 for both models, see Table~\ref{table}.}
\end{figure}

From Eq.~(\ref{eq_density_scaling2}) we see that 
a sensible definition of the onset temperature
becomes 
\be
T_{\rm onset}(\rho) \propto y(\rho)  E_\infty (\rho).
\ee
This is nicely 
illustrated in Fig.~\ref{figure_collapse} (bottom) where $T_{\rm onset}$ 
can be estimated
as the point at which the relaxation data starts to significantly deviate
from the high-$T$ behavior;  the resulting values of $T_{\rm onset}$ are given
in Table~\ref{table}.
The physical significance of the values obtained in
this way are independently confirmed by checking that $T_{\rm onset}$ also 
corresponds to the crossover at which the $T$-dependences of 
the self-diffusion constant, $D_{\rm self}(\rho,T)$, 
and of the relaxation data, $\tau(\rho,T)$, begin to split,  
see Fig.~\ref{figure_decoupling} for an illustration at $\rho=1.2$. 
These values are independently confirmed in Ref.~\cite{Coslovich_preprint}.

The data compiled in Table~\ref{table} actually imply that 
the LJ and WCA models have distinct dynamics in the high-temperature regime 
already since $E_\infty$ of both models only become 
equal at very high density. Both models are also characterized by distinct 
onset temperatures, which differ by a factor 
$\approx 1.6$ at $\rho=1.2$, and, again, slowly
converge to similar values at large density. Finally, 
the qualitative character of the impact of the truncation of 
the potential on 
the slow dynamics is illustrated by the fact that  
the isochoric fragility of the LJ model is independent of density 
while fragility strongly increases with density for the WCA model, 
as parametrized by the evolution of either $x(\rho)$ or $y(\rho)$
which is qualitatively different for both models.

\subsection{Analogy with power-law repulsive spheres}

In the case of the LJ liquid, the density scaling can be expressed, with
essentially as good data collapse as with Eq.~(\ref{eq_density_scaling1}),
by using a power-law density dependence for the scaling parameter (see
Fig.~2c of Ref.~\cite{Berthier-Tarjus09} showing that 
$E_\infty(\rho)  \sim \rho^5$), namely, 
\begin{equation}
\label{eq_density_scaling4} 
\frac{\tau(\rho,T)}{\tau_{\infty}}= {\mathcal F}_2 \left[
\frac{\rho^{\gamma}}{T} \right], 
\end{equation} 
with $\gamma \simeq 5$. In this
form, density scaling was argued to be related to the property that LJ models
are ``strongly correlating liquids''~\cite{DyrePRL&I}, which means that the
fluctuations in the potential energy,
\be
U = \sum_{i < j} v(| {\bf r}_i - {\bf r}_j |), 
\ee
and those in the virial part of the pressure, 
\be
W = - \frac{1}{3} \sum_{i<j} w( | {\bf r}_i - {\bf r}_j | ), 
\ee
with $w(r) = r v'(r)$, 
are correlated at all state points.  The correlation is characterized by a
single parameter, $\Gamma$, defined by a linear fit through a scatter plot 
of the time fluctuations of $U$ and $W$. 
The connection between the density scaling of the
dynamics and the strong-correlation property of the fluctuations shows up in
the approximate relation: 
$\gamma \simeq \Gamma$. 

The rationale which has first been put
forward to justify both
properties~\cite{Coslovich-Roland,DyrePRL&I,Fragiadakis-Roland2010} is as
follows. A dense LJ liquid is dominated by the repulsive forces (the van der
Waals picture of liquids), but the relevant part of the repulsive
interaction involves interatomic distances in the range near to (but less
than) the minimum of the pair potential; in this range, the potential looks
like an effective power law $r^{-3\gamma}$ with an effective repulsion
which is steeper than the 
bare $r^{-12}$ repulsive component of the LJ potential, \textit{i.e.}
$3 \gamma> 12$. In this picture, the LJ model essentially behaves as a
repulsive $r^{-3\gamma}$ power-law interacting model, a model that is known
to rigorously lead to the two considered properties, with indeed 
a strict equality $\gamma = \Gamma$.

The simple rationale given above has however a major flaw: it predicts that
the WCA and the full LJ models should behave in exactly the same way as
their pair potentials coincide for distances shorter than the minimum. This,
as shown above, is clearly wrong. Thus, the WCA-like argument that 
the fluctuations in dense Lennard-Jones
liquids (density scaling and $U-W$ correlation) 
can be understood on the basis of the steep repulsive core only 
is, in general, not valid.

An improvement over the argument was later suggested 
by Dyre and coworkers~\cite{DyreII&III}. They argued that the relevant
interatomic distances actually include the vicinity of the minimum on both
sides of it and that a better description of the LJ potential is through a
power law complemented by a linear term; as the latter has essentially no
effect on the fluctuations at constant density, the fluctuation properties
of the full LJ system are thus predicted to be those of a properly adjusted
power-law repulsive potential. The WCA truncated potential being indeed less
well described by a power law plus a linear term (for distances above the
minimum), this argument could allow one to sidestep the criticism raised
above.

\begin{figure}
\psfig{file=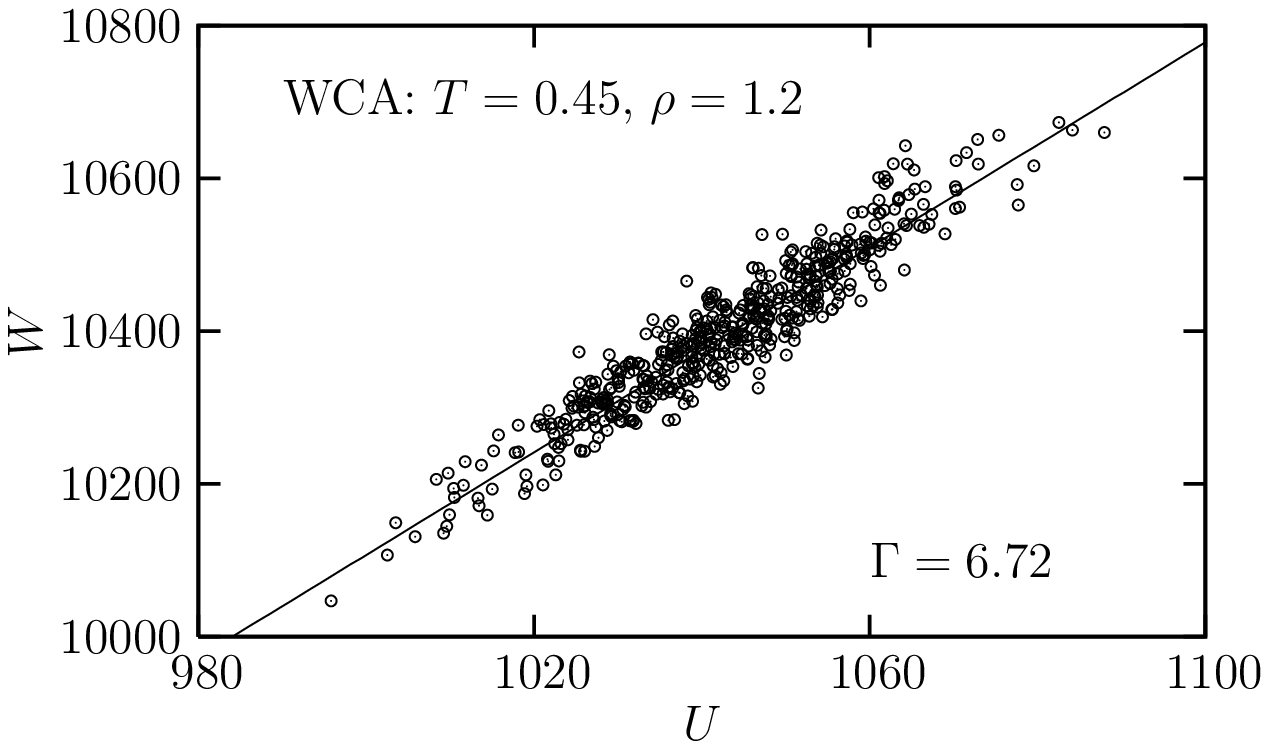,width=8.5cm}
\psfig{file=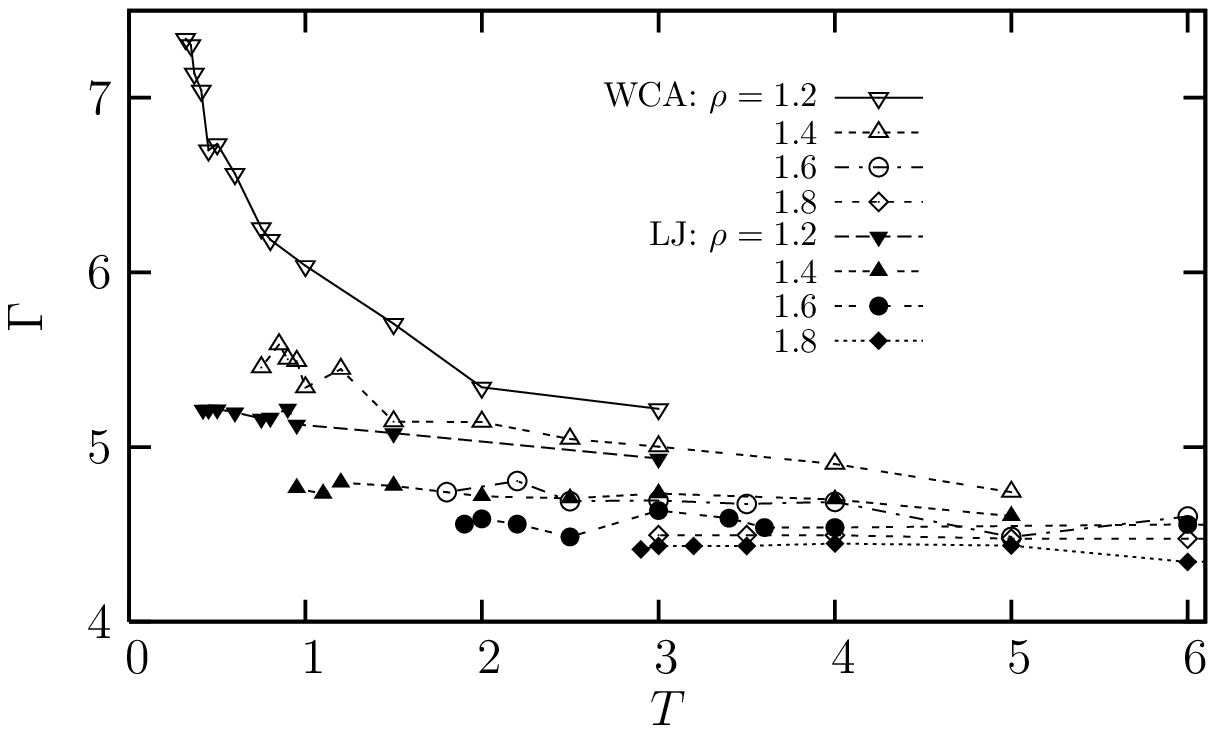,width=8.5cm}
\caption{\label{figure_gamma}
Top: Linear correlation between the fluctuations of the virial $W$ and the
fluctuations of the potential energy $U$ for the WCA liquid at $\rho=1.2$
and $T=0.45$, with a slope $\Gamma$. Bottom: Variation with temperature of
the slope $\Gamma$ of the correlation between the fluctuations of $W$ and
$U$ for the LJ and the WCA liquids at several densities between $1.2$ and
$1.8$.}
\end{figure}

We have repeated the analysis of the correlation between the fluctuations of
the potential energy and those of the virial for the two models. We find
that, as expected from the work of Dyre and coworkers, the LJ model has a
strong correlation at any given density. The WCA
model also has strong correlations between energy and virial at 
any given ($\rho$,$T$) state point~\cite{coslo2}, 
as illustrated in Fig.~\ref{figure_gamma} for $\rho=1.2$ and $T=0.45$. 

The parameter
$\Gamma$ that we extract from the slope of the correlation plot is 
reported in Fig.~\ref{figure_gamma} for a broad range of state points.
It is virtually independent of temperature and  weakly
dependent on density in the LJ case, but has a stronger
density dependence (and temperature dependence when $\rho=1.2$) in the WCA
case (see also Ref.~\cite{Coslovich-Roland}).
Therefore, if one neglects the modest density
dependence of $\Gamma$ for the LJ model (in Fig.~\ref{figure_gamma},
the effective slope $\Gamma$ varies between $4.5$ and $5.2$ over a wide
range of density and temperature), one can indeed conclude that the latter
is ``strongly correlating'' over the whole liquid range with an effective,
constant, value of $\Gamma$ of about $5$. Of course, the data 
in Fig.~\ref{figure_gamma} underlie that there is still some 
arbitrariness in the choice of the value of $\Gamma$, and 
the quality of the density scaling depends on this choice. 

On the other hand, as shown in
Fig.~\ref{figure_gamma}, the variation of $\Gamma$ in the WCA liquid is 
probably too large to be replaced by a single, effective constant. 
In this case, neither the fluctuations nor the dynamics 
are purely determined by reducing the pair
potentials to repulsive power-law ones with an exponent chosen to reproduce
the effective steepness of the potential near the minimum.
Thus, although both phenomena (density scaling and strong-correlation property) seem connected to each other, 
the example of the two models presented in this study shows that 
it is not obvious to know {\it a priori} 
for which materials these observations may apply or not. 

\section{Is the absence or presence of (approximate) scaling due to
attraction or to truncation in the potential ?}
\label{cutoff}

We have seen that the presence (in the LJ model) or the absence (in the WCA
model) of an attractive tail in the pair potentials has a large quantitative
influence on the dynamics and strong consequences for the fluctuations and
the relaxation properties of the liquids. In particular the (approximate)
density scaling observed for the temperature dependence of the relaxation
time in glass-forming liquids and polymers (with the resulting density
independence of the isochoric fragility) is found in the LJ model but not in
the WCA one. A central question then to be raised is whether the differences
stem from the attractive character of the tail \textit{per se} or from the
truncation of the range of the potential to typical interatomic distances. 

A first hint that truncation is the key feature is provided by looking at
the behavior of systems of spheres with purely repulsive power-law
potentials. As already mentioned, such liquids are (rigorously) ``strongly
correlating'' and show an exact density scaling of the relaxation time, as a
consequence of the scale-free power law behavior of the potential.
This example evidences that the presence of attractive interactions 
is not a necessary ingredient in establishing these
properties. 

Additional evidence along the same lines has recently been provided
by Pedersen \textit{et al.}~\cite{Dyre_PRL2010} who showed that the pair
structure \textit{and} the dynamics of the binary LJ mixture at liquid
densities can be very well reproduced by replacing the LJ pair potentials by
\textit{nontruncated} power-law repulsive potentials with an appropriately
adjusted exponent. Note that they chose a repulsion of the 
form $r^{-15.48}$ such that $\gamma = \Gamma = 5.16$, in rough agreement with 
the data shown in previous sections. 

\begin{figure}
\psfig{file=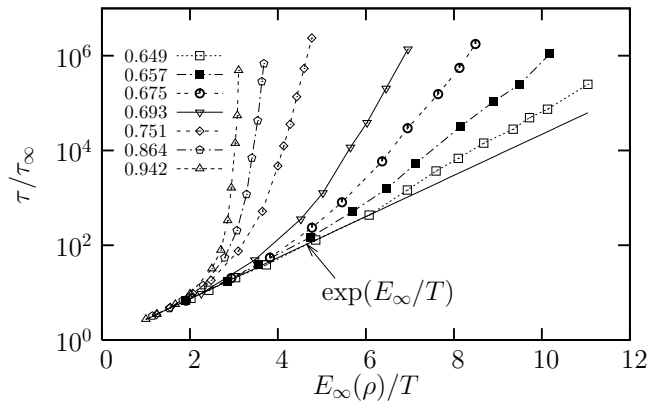,width=8.5cm}
\caption{\label{figure_harmonic} Rescaling of the relaxation time data for
the harmonic repulsive spheres at low temperatures after having normalized the
temperature by the energy parameter $E_{\infty}(\rho)$ chosen to make all
curves collapse at high-$T$. The density $\rho$ is indicated in the figure.
This evidences a dramatic breakdown of density
scaling in harmonic spheres, similar to the one observed for the WCA model.}
\end{figure}

To confirm that truncating the potentials beyond a cutoff of the order of
typical interatomic distances is responsible for the absence of density
scaling seen in the WCA model, we revisit the example of fluid
mixtures of repulsive harmonic spheres, with pair potentials
\begin{equation} 
\label{eq_potential_harmonic} 
\begin{aligned} 
v_{\alpha \beta}(r) &= \frac{\epsilon_{\alpha \beta}}{s}\left(
1-\frac{r}{\sigma_{\alpha \beta}}\right)^{s} , \; {\rm for} \; r \leq
\sigma_{\alpha \beta} \\& =0, \; {\rm for} \; r \geq \sigma_{\alpha \beta},
\end{aligned} 
\end{equation} 
where $s=2$; these systems are commonly used in the context of
zero-temperature jamming phenomena~\cite{OHern-Liu-Nagel}. A wide range of
temperature and density was studied by computer simulation in
Refs.~\cite{berthier-witten1,berthier-witten2}. As noticed in these
references, the isochoric fragility of the system strongly depends on
density. We have replotted in Fig.~\ref{figure_harmonic} the relaxation-time
data in the way described in section IV (and used in Fig. 2 of
Ref.~\cite{Berthier-Tarjus09}) by rescaling the temperature with a
density-dependent energy parameter $E_{\infty}(\rho)$ chosen to make the
high-$T$ data collapse on a single curve. This plot clearly shows that, just
like the WCA model and at odds with the full LJ one, there is no density
scaling of the relaxation for truncated harmonic spheres.

The similarity of behavior between the truncated WCA model and the truncated
harmonic potential on the one hand and between the LJ model and the
repulsive power-law model on the other is a strong indication that the
presence or absence of density scaling in the dynamics of a glass-former
results from the presence or absence of a truncation of all pair
interactions at a typical interatomic distance. This observation 
would perhaps not be very surprising at much lower densities and 
temperatures (see Fig.~\ref{figure_phase_diagramLJ}) since,
{\it by construction}, the WCA potential coincides with the one of harmonic 
spheres in Eq.~(\ref{eq_potential_harmonic}) near the cutoff. The surprising
feature is that this analogy seems to be relevant up to the 
liquid densities studied in the present work, and it results  
in the WCA model behaving differently from the LJ system.
As discussed in the
following section, this conclusion casts doubts on a description of
supercooled liquids in terms of the jamming scenario.

\section{Glass transition versus jamming phenomenon}
\label{jamming}

The jamming paradigm~\cite{Liu-Nagel-et-al,Liu-Nagel} 
has been put forward to bring together
in a common picture a wide breadth of phenomena and systems involving
sluggish dynamics and freezing in an amorphous state. A step to go beyond
qualitative comparisons has been taken with the proposal that the slowing
down of all jamming systems, whether driven by temperature, density or
applied force, is controlled by a zero temperature and zero applied force
critical point, ``point J''~\cite{OHern-Liu-Nagel}. This proposal has been
criticized on several grounds, in particular concerning the uniqueness of
point J itself~\cite{many1,many2}, and the possibility that a distinct
glass critical point, which could be dubbed ``point G'', 
controls the finite temperature dynamics of jamming  
systems~\cite{Zamponi-Parisi}.  
Here, we focus on the ability of the jamming scenario
and of the associated point J (whether unique or not) to describe the
slowdown of relaxation of actual glass-forming \textit{liquids} (and
polymers).

A transition near point J only takes place at zero temperature and moderate 
density, and it can only be rigorously defined for interaction 
potentials that are zero beyond some distance and repulsive within it.
Therefore,  to be applicable to liquids in their (experimentally) 
accessible range, the
jamming scenario must assume that the role of the attractive tail and more
generally of longer-ranged (longer than the typical interatomic distance)
interactions is negligible in the physics of the slowing down. 
To extrapolate the behavior of a real
liquid to such conditions, one must then get rid of the attractive forces,
which give rise to the gas-liquid coexistence curve and prevent one from
taking the liquid to zero temperature below some density (see the phase
diagram in Fig.~\ref{figure_phase_diagramLJ}), and eliminate the pair
interactions beyond some cutoff that physically determines the typical
inter-particle distance. Note that this is precisely what the WCA procedure
achieves. However, as we have seen above, the behavior of the WCA model
strongly deviates, both qualitatively and quantitatively, from that of the LJ model and more generally that of  realistic liquids, 
except at very high densities that are beyond any
experimentally accessible range. It is this observation which 
leads to question the ability of the jamming scenario to describe 
the dynamics of viscous liquids.

Actually, systems of particles interacting through truncated repulsive
interactions, such as the harmonic-like models in 
Eq.~(\ref{eq_potential_harmonic}) or the WCA model, behave as
``effective'' hard-sphere systems at low temperature and low
pressure. In such thermodynamic conditions, 
it is the behavior of the pair potential near the cutoff which governs
the physics, and if temperature is low, soft repulsive spheres can be seen 
as ``disguised'' hard spheres~\cite{berthier-witten1,berthier-witten2}. 
As shown in Fig.~\ref{figure_harmonic} and discussed in detail
in Ref.~\cite{berthier-witten1} this directly implies that the isochoric
fragility of such liquids strongly depends on density. This is 
a first qualitative distinction between the physics near the jamming 
transition and the one of supercooled liquids. 

\begin{figure}
\psfig{file=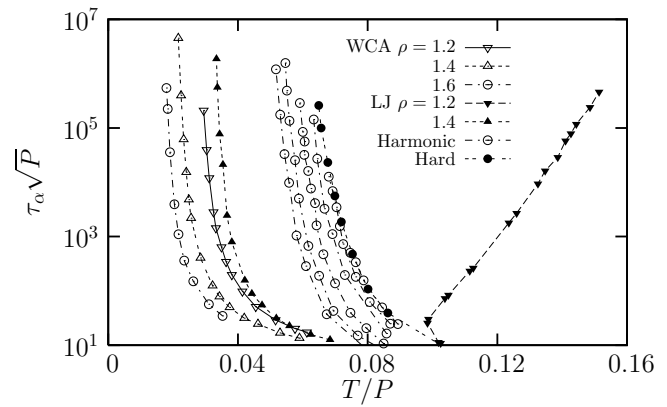,width=8.5cm}
\caption{\label{figure_haxton} Hard-sphere-like scaling of the relaxation time. We use 
Eq.~(\ref{eq_hardsphere_scaling}) and follow the evolution 
of the relaxation dynamics from hard spheres~\cite{gio}, 
to harmonic spheres~\cite{berthier-witten1}, to WCA and finally LJ systems,
which exhibit increasingly larger deviations from hard-sphere behavior.}
\end{figure}

A second consequence, discussed in Ref.~\cite{Xu-Liu-Nagel}, 
is that relaxation data for soft spheres with truncated repulsive interactions at different
state points converge in the low $T$ and low $P$ limit to a 
hard-sphere-like scaling, such that~\cite{Xu-Liu-Nagel},
\begin{equation}
\label{eq_hardsphere_scaling}
\frac{\tau(\rho,T)}{\tau_{\infty}}= {\mathcal H} \left[ 
\frac{P(\rho,T)}{T} \right].
\end{equation}
where $\tau_{\infty}$ has a residual (trivial) dependence on either $T$ or $P$
(\textit{e.g.} in $1/\sqrt{T}$ or $\sqrt{P}$) 
that it is now important to take into
account as the range of pressure and temperature spanned is very large
[compared to the liquid range considered for
Eqs.~(\ref{eq_density_scaling1}-\ref{eq_density_scaling4})]. 
The domain of application of this scaling is demonstrated in 
Fig.~\ref{figure_haxton} where we include data obtained from 
simulations of hard spheres taken from Ref.~\cite{gio}. The harmonic-sphere data of Fig.~\ref{figure_harmonic}  
indeed converge to the hard-sphere behavior when density is not 
too large, in agreement with the results of Ref.~\cite{Xu-Liu-Nagel}, 
and display small deviations 
from Eq.~(\ref{eq_hardsphere_scaling}) when density is larger. These 
deviations were empirically addressed in 
Ref.~\cite{berthier-witten1,berthier-witten2} using a $(\rho,T)$-dependent
``effective'' volume fraction. 
Dynamics in such systems with truncated
repulsive potentials at low temperature and low-to-moderate pressure seems
then dominated by (renormalized) free-volume or congestion effects, as for a hard-sphere
fluid.  
Whether (or not) the slowdown of relaxation is controlled by a $T=0$
point J~\cite{OHern-Liu-Nagel} or by a $T>0$ point
G~\cite{berthier-witten1,Zamponi-Parisi}, the jamming and glass transitions
appear in any case strongly intertwined in this case.

However, when moving from harmonic to WCA potentials in 
Fig.~\ref{figure_haxton}, we observe much stronger deviations 
from the hard-sphere scaling, and the data in fact converge 
to the vertical axis $T/P = 0$ when density increases. 
This is expected, as in the large density limit we expect 
the relaxation-time data to be driven by the scaled variable 
$T^{5/4}/P$, which becomes exact for the power law $r^{-12}$ repulsion.

Making the final step from WCA to LJ models in Fig.~\ref{figure_haxton}
we realize that the effect of the attractive forces becomes dramatic
for normal liquid densities. Since attractive forces induce the presence
of the gas-liquid phase separation, the pressure near the coexistence curve  
drops to zero when temperature decreases 
much more rapidly than in the absence of the attractive tails. The effect
is indeed spectacular for the density $\rho=1.2$ in Fig.~\ref{figure_haxton}
since the LJ data have a qualitative behavior opposite to that 
of truncated purely repulsive systems. This echoes Voigtmann's 
remarks~\cite{Voigtmann}
emphasizing 
that WCA and LJ models appear more similar when using $(\rho,T)$
than $(P,T)$ variables, for the same reason. This observation 
is experimentally relevant as atmospheric-pressure data for viscous liquids
virtually all fall in this regime~\cite{Voigtmann}, implying that 
the hard-sphere-like 
scaling in Eq.~(\ref{eq_hardsphere_scaling}) is not obeyed for liquids
at normal conditions. 

This second argument confirms our claim that in the range of
pressure, density, and temperature that corresponds to the actual
liquid/supercooled liquid range, the glass transition of the liquid and that
of jamming models described by truncated repulsive potentials are different
and cannot be controlled by the same $T=0$ point J critical point.

We finally make the aside that the models defined through truncated soft
repulsive potentials differ when considered at high temperature and density.
The (unbounded) WCA potential leads to a system that converges to the full
LJ model, as shown above; asymptotically, at large $\rho$ and $T$, the WCA
and the LJ models behave as a rescaled hard-sphere system (the rescaling is
however not the same as at low pressure and temperature).  On the contrary,
the (bounded) harmonic-like models rapidly reach very high densities at
which large numbers of atoms overlap, which seems to confer to these models
some sort of mean-field behavior~\cite{Myasaki}, with peculiar 
physical consequences~\cite{moreno}.

\section{Conclusion}
\label{conclusion}

Through an extensive comparison of the behavior of a standard Lennard-Jones
glass-forming liquid and that of its WCA reduction to a model with truncated
pair potentials without attractive tails, we have shown that the slowdown of
relaxation is quantitatively and qualitatively different in the two models
while the equilibrium pair structure remains very similar. The differences
in the dynamics of the two models decrease as density increases but one has
to reach unphysically large values of the density for seeing a full
convergence of behavior. Clearly, the presence or absence of the attractive
tails cannot be neglected in the viscous regime, where a liquid can no
longer be simply described according to the conventional van der Waals
picture: fluctuations associated with the tails of the interaction
potentials do play a role. 

At this point, it should be stressed that, in spite of the observed
differences, the purely repulsive WCA model is a glass-former, no less than
the full LJ one. The attractive tail is not a necessary ingredient to
trigger a slowdown of relaxation. The WCA model displays features
generically associated with glass formation. For example, as seen from
Fig.~\ref{figure_self_intermediate_scattering}, the associated time
dependence of the self intermediate scattering function is characterized by
a stretching and the appearance of a plateau as one lowers the temperature.
Moreover, the dynamics is increasingly heterogeneous~\cite{chandlerreview} and
shows a decoupling between diffusion and structural relaxation, 
see Fig.~\ref{figure_decoupling}.
Not surprisingly, the temperature dependence of the relaxation time data can
also be fitted by the same functional forms, \textit{e.g.} the
Vogel-Fulcher-Tammann formula or a B\"assler-type
expression~\cite{Elmatad-Chandler}, for both models. This is enough to invoke
analogies, but not necessarily to assign the slowing down to the same
mechanisms.

The nonperturbative role of attractive forces 
may be tentatively attributed to the growth,
as temperature decreases, of some form of heterogeneities in the liquid that
are not captured by the static pair density correlations. We have not
investigated subtler characterizations of the liquid structure,
\textit{e.g.} via higher-order correlation functions or topological
measures, but preliminary studies seem to indicate that these are more
sensitive to the attractive tails than the pair density correlations and
show structural differences between LJ and WCA models that grow as
temperature is lowered~\cite{royall,Coslovich_preprint,davepts}. 
This is certainly a
worthwhile line of research to pursue.

In trying to characterize the physical significance of the increasing
quantitative difference in the dynamics of the full LJ model and of the
truncated repulsive WCA one, we have stressed that this difference could not
be accounted for by a mere one-parameter rescaling of the data. (The fact
that the difference cannot be reproduced either by a large amplification in
the dynamics of the small differences in the static pair density
correlations is considered
elsewhere~\cite{Berthier-TarjusMCT,Berthier-Tarjus2}.) The temperature-driven
slowdown of relaxation in the two models actually show
\textit{qualitatively} distinct behavior, as manifested by the absence in
the WCA model of the density scaling that is found in the full LJ model and
in experimentally studied glass-forming liquids and polymers. Moreover, in
the LJ and WCA models, this feature appears to be related to the properties
of the correlation between the fluctuations of the virial and of the
potential energy.

Still concerning the physical significance of our results, we have addressed
the question of whether the observed differences between the WCA and the LJ
models in the viscous regime are due to the attractive nature of the missing
tail (in the WCA model) or to the fact that the tail is truncated at a
cutoff corresponding to typical interatomic distances (in the WCA model). We
have provided strong evidence that the key feature is the truncation of the
interaction potentials and we have stressed that models with truncated
repulsive potentials show qualitatively distinct temperature-driven slowing
down than glass-forming liquids. This casts serious doubts on the
possibility of describing glass formation in liquids by a jamming scenario
in which the slowdown of relaxation is controlled by a critical point at
zero temperature and density less than liquid densities.

\acknowledgments
We thank D. Coslovich, D. Reichman and C. P. Royall 
for fruitful exchanges about this work.
L. B.'s work is partially funded by R\'egion Languedoc-Roussillon.

\end{document}